# Deep Learning Techniques for Cervical Cancer Diagnosis based on Pathology and Colposcopy Images

Hana Ahmadzadeh Sarhangi[1], Dorsa Beigifard[1], Elahe Farmani[2], Hamidreza Bolhasani[1+]

*Abstract*— **Cervical cancer is a prevalent disease affecting millions of women worldwide every year. It requires significant attention, as early detection during the precancerous stage provides an opportunity for a cure. The screening and diagnosis of cervical cancer rely on cytology and colposcopy methods. Deep learning, a promising technology in computer vision, has emerged as a potential solution to improve the accuracy and efficiency of cervical cancer screening compared to traditional clinical inspection methods that are prone to human error. This review article discusses cervical cancer and its screening processes, followed by the Deep Learning training process and the classification, segmentation, and detection tasks for cervical cancer diagnosis. Additionally, we explored the most common public datasets used in both cytology and colposcopy and highlighted the popular and most utilized architectures that researchers have applied to both cytology and colposcopy. We reviewed 24 selected practical papers in this study and summarized them. This article highlights the remarkable efficiency in enhancing the precision and speed of cervical cancer analysis by Deep Learning, bringing us closer to early diagnosis and saving lives.**

[1] Department of Computer Engineering, Islamic Azad University Science and Research Branch, Tehran, Iran.
[2] Tehran University of Medical Sciences, Tehran, Iran.
[1+] Corresponding Author: hamidreza.bolhasani@srbiau.ac.ir

# 1. Introduction

## 1.1. Cervical Cancer, Screening, and Diagnosis

In women's health, cervical cancer has become a prevalent issue, ranking as the fourth most commonly diagnosed cancer. To put things in perspective, in 2020, around 604,000 new cases of cervical cancer were identified worldwide [1]. Cervical cancer, as a whole, is the fourth leading cause of death globally, responsible for the loss of 342,000 lives. Unfortunately, the majority of these fatalities occur in sub-Saharan Africa, Melanesia, South America, and Southeast Asia [1]. However, despite the concerning statistics, it's essential to recognize that cervical cancer is considered nearly entirely preventable and treatable if diagnosed in the early stages.

Cervical cancer is a severe medical condition mainly resulting from the Human papillomavirus (HPV) [2]. The majority of cervical cancer cases can be attributed to this virus, and is highly prevalent in the general population. Different types of HPV are often transmitted sexually or through other skin-to-skin contact. Of the various types of HPV, HPV-16 and HPV-18 are considered high-risk and the most common causes of advanced lesions and cancer [3]. These types of HPV are known to cause abnormal cell growth in the cervix, and the growth of cancerous tumors can be a possible outcome in cases where this condition is present. It is crucial that individuals, especially women, take the necessary precautions to prevent and detect cervical cancer early on, such as regular screenings and vaccinations.

In the process of cervical cancer development, irregular alterations in the squamous cells, labeled as Cervical Intraepithelial Neoplasia (CIN), are strongly connected to infection by HPV types that pose a higher risk. As stated by the World Health Organization (WHO), there are two types of CIN: CIN 1, which is low grade, and CIN 2 or 3, which are high grade (Fig 1). While not all cases of CIN progress to cervical cancer, untreated high-grade CIN can eventually lead to invasive cervical cancer over several years. Additionally, CIN 1 can be categorized as LSIL (Low-grade Squamous Intraepithelial Lesion), and CIN 2 and CIN 3 can be categorized as HSIL (High-grade Squamous Intraepithelial Lesion). Cervical cancer can often be detected through various screening tests, including Visual Inspection of Acetic Acid (VIA), Cytology (pap smears and liquid-based cytology), HPV testing, and Colposcopy [4]. These tests are designed to identify CIN in the cervix of apparently healthy asymptomatic women, which can be a sign of precancerous lesions or other potential risks for cervical cancer. Regular testing is important for preventing and treating cervical cancer, as early detection can significantly improve treatment outcomes. The steps involved in cervical screening and diagnosis are shown in Fig 2.

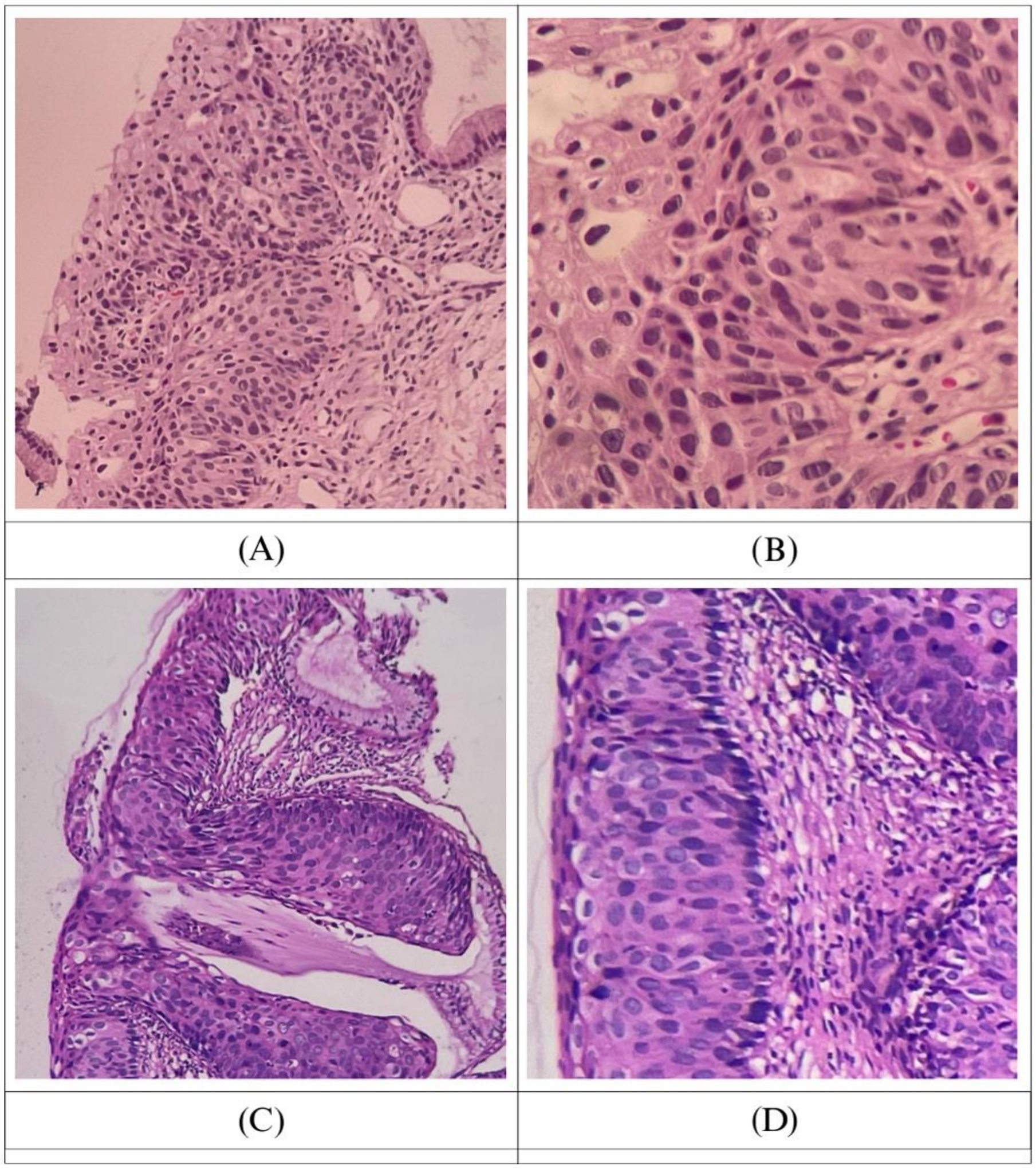

Fig. 1. Low-grade CIN (A, B) and High-grade CIN (C, D)

VIA is a quick and effective way to screen for cervical cancer, where 3-5% acetic acid is used for the cervix to turn pre-cancerous cells white for visual identification by a healthcare provider. The healthcare provider will then perform an examination with the unaided eye or a colposcope for a visual evaluation and identify any suspicious areas that require a closer look. VIA is accessible and relatively inexpensive in resource-limited settings and can be carried out by healthcare professionals at the mid-level, such as nurses and midwives. However, its sensitivity is lower than Cytology and Colposcopy. Cytology can be performed in 2 ways: as a pap smear or liquid-based cytology test. The procedure for performing a Pap smear involves gathering cervical cells and transferring them onto a microscope slide for analysis. Liquid-based cytology is a test similar to a pap smear that collects cells from the cervix and observes them under a microscope but uses liquid to preserve cells before viewing. This modification in the testing process aims to increase the test's accuracy and sensitivity [5].

HPV testing is typically done by collecting samples from the cervix with a small brush and spatula and sending the sample to the laboratory. In a laboratory, HPV testing is typically done using a process called Polymerase Chain Reaction (PCR) or hybridization [6]. PCR is a laboratory technique that can detect and amplify specific sequences of DNA, including the DNA of HPV. Hybridization is a process in which the DNA of HPV binds to specific probes that have been coated on a filter or microchip [6]. If the screen test is positive for CIN, confirming the diagnosis involves undergoing a colposcopy followed by directed biopsies. It is typically performed in a healthcare setting by a gynecologist or other qualified provider using a colposcope, which is a small, specialized microscope that magnifies the vaginal area and can be used to visualize abnormal tissue. A biopsy may be sent to a laboratory for examination and testing. A pathologist will examine the tissue under a microscope and look for any signs of abnormal cells, including cancer cells.

The WHO advises women to start getting screened for cervical cancer at the age of 30 and follow a regular screening schedule of 5 to 10 years after that. [7]. However, the majority of cancer screening programs have only occurred exclusively in countries with high levels of income. In low-middle-income countries, only a minority of women, roughly 20%, have undergone testing for cervical cancer, while in high-income countries, more than 60% of women have been screened for this disease.[8]. This could be due to a shortage of expertise or the high costs of screening.

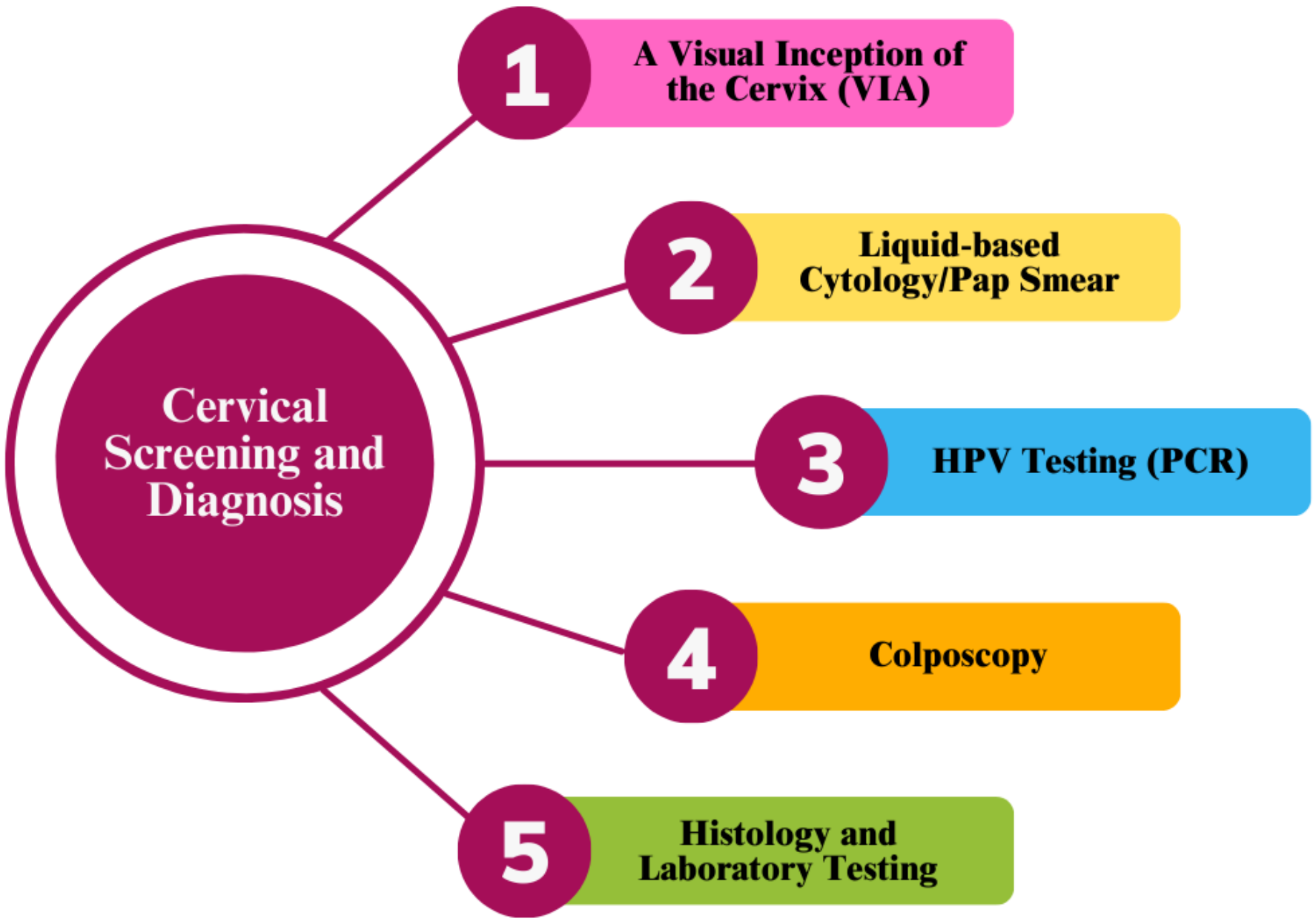

Fig. 2. Steps involved in cervical screening and diagnosis

### 1.2 Deep Learning and Machine Learning

Artificial intelligence (AI) is a study that combines machines and human problem-solving approaches. It consists of Machine Learning (ML), Natural Language Processing (NLP), robotics, and expert systems [9]. Machine learning and deep learning (DL) stand out as the primary categories within the broad field of artificial intelligence. Arthur Samuel initially defined machine learning in 1959 [10]. He portrayed it as “A field of study that gives computers the ability to learn without being explicitly programmed.”. The popularity of ML is due to its ability to recognize patterns and make decisions without being explicitly programmed. Numerous types of algorithms exist that aim to acquire knowledge from data and make predictions or formulate choices. Humans have been looking for ways to enable computers to learn and process information like the human brain [11]. As a result, Artificial Neural Networks (ANN) are inspired by the human learning system. Humans learn by observing and comprehending the world through trial and error. Therefore, ANN is updated to improve its accuracy as new information is acquired.

Deep learning is a subset of ML that emphasizes the training and development of ANNs with multiple layers. A pivotal breakthrough in deep learning occurred when Yann LeCun introduced a novel architecture known as Convolutional Neural Networks (CNNs) [12]. CNN brought about a revolution in the field by enabling the creation of deeper neural network models with multiple hidden layers. The effectiveness of CNN improved when Alex Krizhevsky developed a groundbreaking architecture (AlexNet) using CNNs, showcasing the potential of deep learning in computer vision [13]. Following these advancements, there was a tremendous surge in practical applications of deep learning in the medical field [14-17], as shown in Fig 3. The capacity of DL to identify complex patterns and aid early detection and treatment contributes to cancer research and diagnosis.

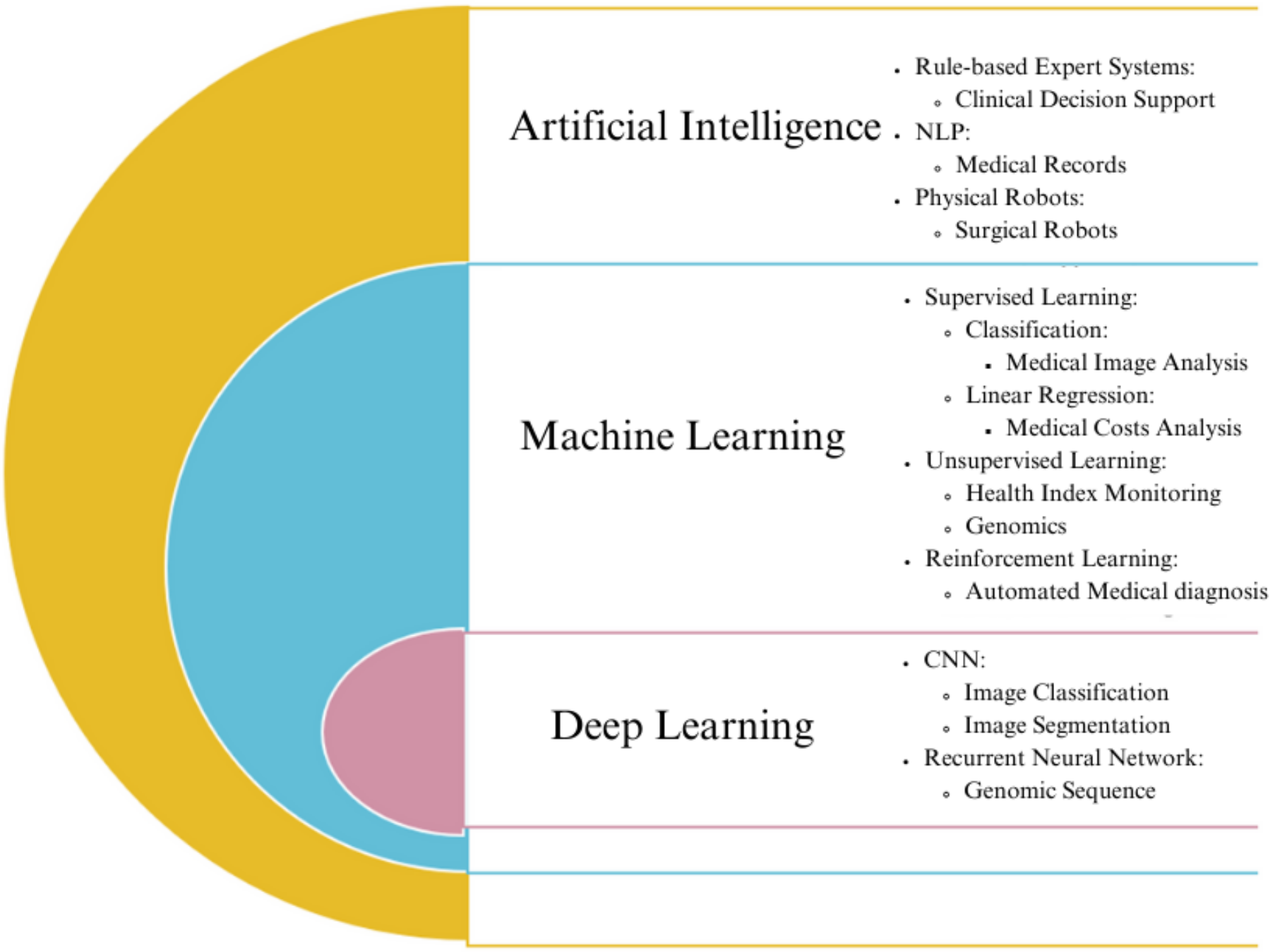

Fig. 3. The role of AI, ML, and DL in medicine

Neural networks have been applied in medicine since 1960 [17]. The progress of advancements in deep learning is attributed to key factors. First, the accessibility of diverse medical data, including MRI, CT scans, genomics, etc. Second, the evolution of deep learning architectures, and third, the introduction of potent hardware like GPU and TPU [18]. These factors have made a transformative phase in medical AI. Greener et al. [19] conducted a comprehensive review of machine learning algorithms centered on the application of DL in biology and medicine. The study provides practical examples of the applications of these algorithms and recommends CNN as the optimal choice for image analysis.

In addition, digital pathology has revolutionized the field of pathology by allowing the storage and remote access of digital data from glass slides through brightfield and fluorescent slide scanners. Whole-Slide Imaging (WSI) scanners convert histopathology, immunohistochemistry, or cytology slides into digital images, which can be interpreted, managed, and analyzed using AI algorithms. Computer-Aided Diagnosis (CAD) has played a pivotal role in medical screening since the 1960s. With the advancements in AI techniques and machine learning, CAD systems, and digital pathology have become even more sophisticated. These technologies aid doctors in making more precise diagnoses, leading to improved patient outcomes, and are now integral to the medical industry.

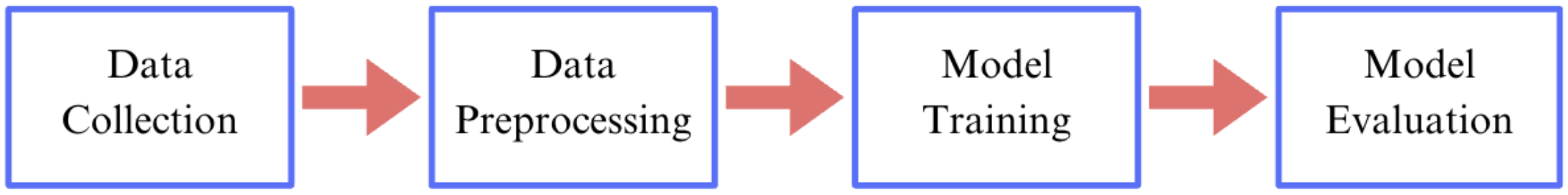

Fig. 4. Process of a deep learning approach

A deep learning approach comprises four primary processing steps, illustrated in Fig 4, frequently encountered in research papers. Data collection stands as the initial and pivotal phase in the learning process. Deep learning models benefit from larger datasets, and a lack of sufficient data can hinder algorithm performance. The scarcity of large datasets is particularly pronounced in medical imaging, especially for cytology and colposcopy images. However, this challenge can be addressed through techniques such as data augmentation, which involves applying transformations to images to generate additional examples, and by using Generative Adversarial Networks (GAN) technology to create new image instances. The next step in the process involves dataset preprocessing, where the dataset is split into three segments: the training set, validation set, and test set. During this stage, normalization and standardization can proceed to prepare the data for improved training and enhanced performance. Normalization entails scaling the features to a consistent size, making them suitable for input into the deep learning model. Additionally, another technique used in this phase to enhance the representation of cellular structure and properties is staining, such as the Pap or H&E method [20].

The training phase encompasses a range of intricate tasks, including classification, segmentation, and detection. For instance, in cytology, the classification tasks can be categorized into two main groups. One of these categories focuses on cell-level classification, where the objective is to categorize individual cell images. Initially, Regions of Interest (ROIs) are obtained from Whole-Slide Images (WSIs) and then utilized to train Deep Neural Network (DNN) models. These models subsequently generate classifications for cervical cell types. In parallel, another classification approach is slide-level classification, which is dedicated to assessing malignancy risk grading within WSIs. Segmentation task plays a leading role in cervical cancer diagnosis. It involves the process of partitioning an image into distinct regions, where each segment comprises pixels or elements that share common characteristics or belong to the same class. In this field of study, segmentation finds numerous applications, such as segmenting cell nuclei or cytoplasm using techniques like image histogram thresholding, optical density measurements, and image gradient analysis. Another important application of segmentation is to identify and distinguish cancerous from non-cancerous regions within cervical tissue samples [21]. Furthermore, automatic contour segmentation techniques are applied to separate cellular ROI from the background, bolstering the precision of the analysis [22]. Identifying particular objects within an image is known as object detection. In the context of medical images, this technology is employed to pinpoint anomalies, such as lesions, and categorize a wide range of objects [23]. It is capable of discerning irregular cells within multi-cell images and assigning them to various cytological categories. Additionally, this method is adept at addressing challenges related to overlapping and variations in the sizes of target cells.

Assessing the performance of models, evaluation metrics differ across the domains of classification, segmentation, and object detection, with each type of task having its specific set of metrics. These metrics are designed to evaluate the efficiency of algorithms in distinct ways. In the context of classifying pathological tumors, essential evaluation parameters include TP (True Positive), FP (False Positive), TN (True Negative), and FN (False Negative). Below, explanations are provided for these terms specifically concerning cancer screening. Furthermore, most classification performance metrics, such as sensitivity, specificity, and accuracy, rely on these parameters, and their formulas are provided in Table 1. Segmenting cancerous cells from normal ones requires different evaluation metrics. Qualifying whether the model correctly partitions input images at the pixel or object level necessitates distinct metrics focused on identifying object boundaries and fine-grained details. Table 2 includes the segmentation metrics as well.

- TP, or True Positive, indicates a correct prediction by the model when the condition is a cancerous cervical lesion.
- TN, or True Negative, represents a correct prediction by the model when the condition is non-cancerous.
- FP, or False Positive, signifies incorrect predictions where the model falsely identifies the condition as cancerous when it is non-cancerous.
- FN, or False Negative, also signifies incorrect predictions, but in this case, it means the model falsely predicts a non-cancerous condition when it is cancerous, potentially leading to critical issues.

Table.1. Evaluation metrics of classification and segmentation

| **Classification** | | **Segmentation** | |
|---|---|---|---|
| **Metrics** | **Formulas** | **Metrics** | **Formulas** |
| Precision (PREC) | $\frac{TP}{TP+FP}$ | Dice Similarity Co-efficient (DSC) (A,B) | DSC (A,B) = 2(A∩B)/(A+B) |
| Recall (REC), Sensitivity (SEN) | $\frac{TP}{TP+FN}$ | Intersection over Union (IoU) or Jaccard index | J (A, B) =$\frac{\|A \cap B\|}{\|A \cup B\|}$ |
| F1 score, Dice Coefficient (DICE) | $2\times\frac{Preciosn\times Recall}{Precision+Recall}$ | Zijdenbos similarity index (ZSI) | ZSI=$\frac{2TP}{2TP+FP+FN}$ |
| Specificity (SPEC) | $\frac{TN}{TN+FP}$ | Mean Average Precision (mAP) | mAP = $\frac{1}{N}\sum_{i=1}^{N} AP_i$ |
| Accuracy (ACC) | $\frac{TP+TN}{TP+TN+FP+FN}$ | Average Recall (AR) | AR = $\frac{1}{n}\sum_{t=0}^{n-1} AR_t$ |

## 2. Public Datasets for Cytology and Colposcopy Images:

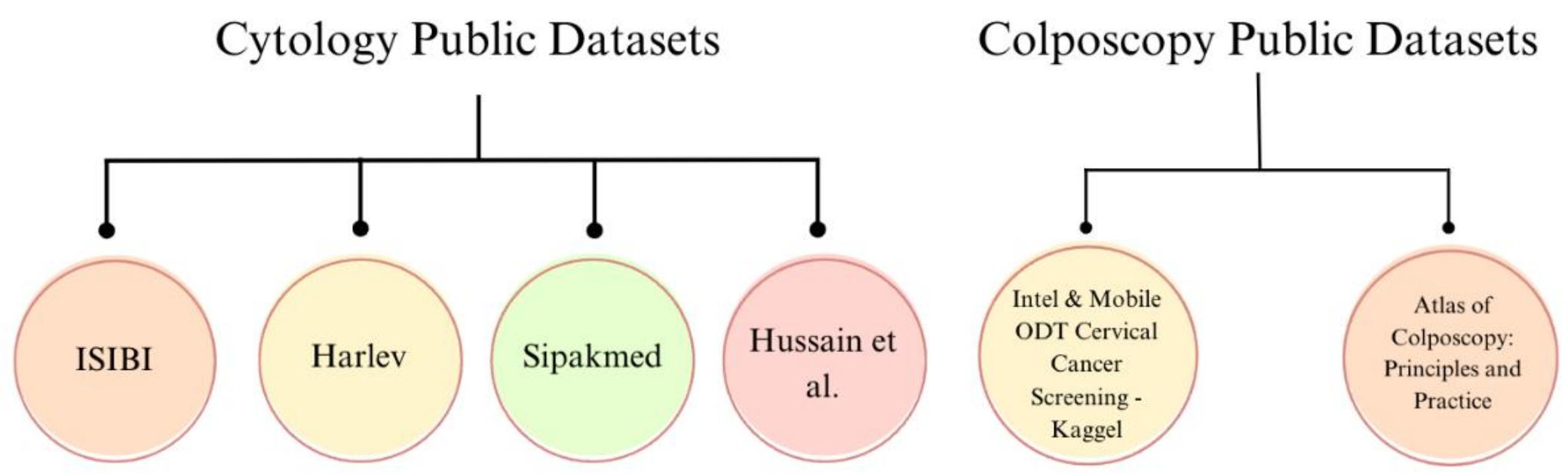

Fig. 5. Public datasets for cytology and colposcopy images

This section aims to review public datasets for cytology and colposcopy gathered from studies mentioned later in this article, which are listed in Fig 5.

### 2.1 Cytology:

Harlev: The Pap smear benchmark database has been a prominent resource in numerous research endeavors involving the application of deep learning techniques since 2005. This dataset comprises individual cell images annotated with ground truth data, encompassing a total of 917 cells, with 675 identified as abnormal and 248 as normal. The specimen preparation procedures adhere to the established protocols of traditional Pap smear and Pap staining techniques, ensuring the reliability and consistency of the dataset. This most used database originates from Harlev University Hospital in Denmark. It has been categorized into seven significant classes: classes (a) to (c) denote healthy cells, while classes (d) to (g) pertain to abnormal cells, including (a) superficial squamous epithelia, (b) intermediate squamous epithelia, (c) columnar epithelia, (d) mild squamous non-keratinizing dysplasia, (e) moderate dysplasia, (f) severe dysplasia, and (g) carcinoma in situ [24].

SIPAKMED: In 2018, X constructed a pap smear image dataset was created, which includes 5 distinct categories. The first two categories represent normal cells, the next two denote abnormal cells, and the final category comprises cells from the transformation zone. These categories are (a) Superficial-Intermediate, (b) Parabasal, (c) Koilocytotic, (d) Dyskeratotic, and (e) Metaplastic cells. This dataset consists of 966 Pap smear slides, totaling 4,049 labeled single-cell images. To provide more specific details, there are 2,411 cytology-negative images and 1,638 cytology-positive images [25].

ISBI: Another dataset commonly used in research is the ISBI 2014 and 2015 challenge dataset, primarily employed for overlapping cell segmentation purposes. This dataset consists of cervical cell images paired with corresponding ground truth information. The 2014 dataset comprises 16 real cervical cytology EDF images and 945 synthetic images. The number of overlapping cells in the 2014 dataset varies from 2 to 5. In the subsequent ISBI 2015, Nine sets of actual cervical cytology EDF images are provided, each with its corresponding volume images. The number of overlapping cells in the ISBI 2015 dataset ranges from 2 to 10.

Hussain et al. [26]: It is an LCB dataset conducted in India, comprising a total of 963 images. Among these, 613 are classified as normal cells, and 350 fall into the abnormal category, further divided into three classes: 113 LSIL images, 163 HSIL images, and 74 SCC images, as per the TBS classification. This dataset is publicly available.

### 2.2 Colposcopy

In the field of colposcopy, public datasets are scarce, and most research studies rely on private datasets. There are only two available datasets, one from Kaggle community developers [27] called "Intel & Mobile ODT Cervical Cancer Screening" and the other from the International Agency for Research on Cancer (WHO) called "Atlas of Colposcopy: Principles and Practice" [28]. The "Intel & Mobile ODT Cervical Cancer Screening" dataset consists of 1481 cervix images categorized into two groups based on their visual appearance: normal (non-cancerous) and abnormal (accessed in 2021). The "Atlas of Colposcopy" case studies include images of high-grade, low-grade, and normal colposcopy. The dataset includes images of the green filter, Lugol's iodine, and aceto-white effect for each patient. The high-grade and normal colposcopy images are included in the dataset (accessed in 2022).

## 3. The Most Popular Deep Learning Architectures Applied in Cytology and Colposcopy

Many of the papers we reviewed have employed ImageNet for pre-training CNNs with various architectures. Some of the most widely used architectures in cervical cancer diagnosis include VGG-16, ResNet50-101, Inception-v3, and EfficientNetB0-3. In this section, we will provide brief descriptions of these network architectures, as shown in Fig 6 and Fig 7.

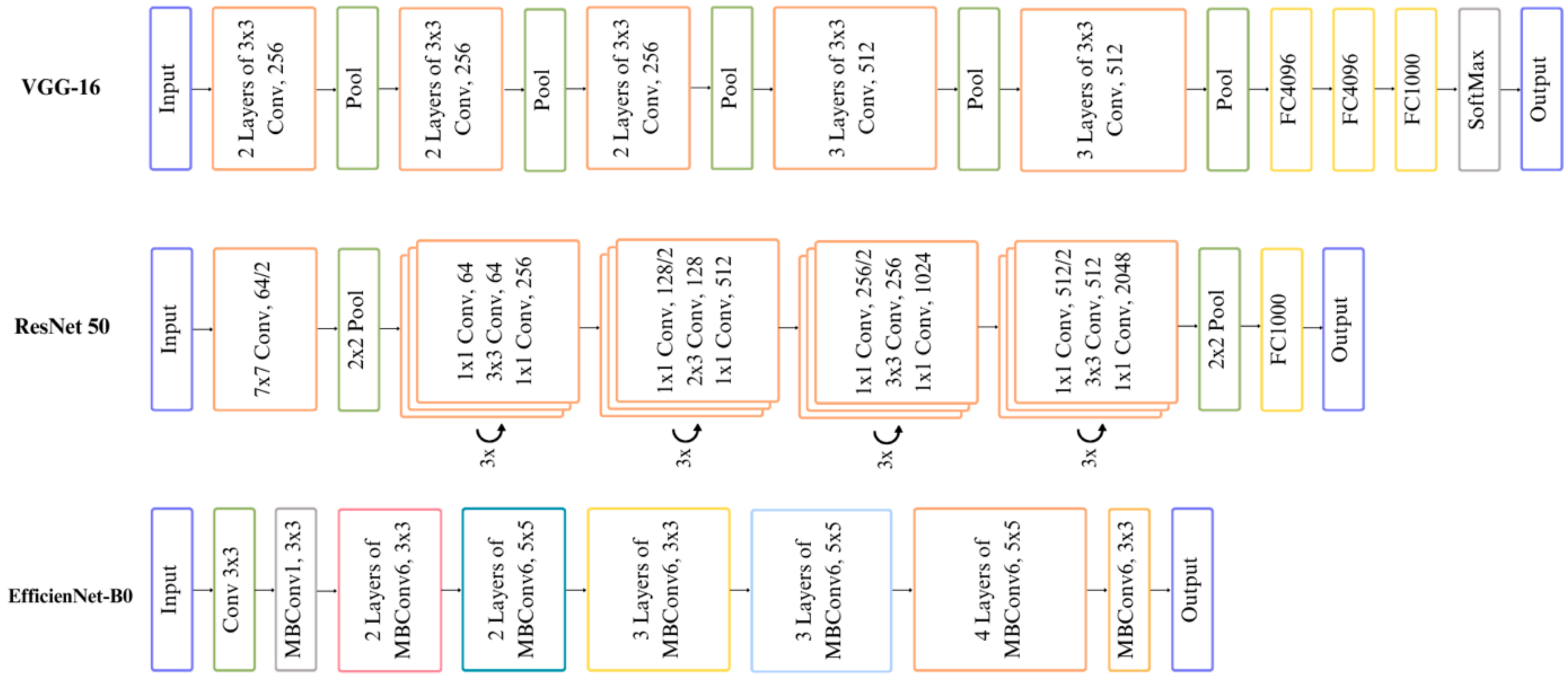

Fig. 6. The most popular deep learning architectures applied in cytology and colposcopy

1) VGG-16

The VGGNet, introduced by Simonyan and Zisserman from the University of Oxford, achieved second place in the 2014 ILSVRC competition. VGGNet offers different versions, ranging from 11 to 19 layers, with its primary focus being on network depth. It takes as input an RGB image of size 224x224 pixels and employs 3x3 filters for convolution. Additionally, 1x1 filters are applied before the ReLU activation function to enhance network performance while preserving spatial information with a stride of 1. This architecture includes three fully connected layers, where the initial two layers consist of 4096 channels, and the last layer comprises 1000 channels, each representing a distinct class. VGGNet incorporates five 2x2 Max pooling layers with a stride of 2, and it concludes with a softmax layer. The term 'VGGNet-16' indicates that the network has 16 layers, as shown in fig 6. This network architecture is widely recognized and widely adopted in the field.

2) ResNet

A Deep Convolutional Neural Network (DCNN) known as ResNet took first place in the ImageNet Large Scale Visual Recognition Challenge (ILSVRC) competition. Prior to the introduction of ResNet, training neural networks with a substantial number of layers posed a significant challenge due to the issue of vanishing gradients. ResNet, short for Residual Network, successfully addressed this problem by introducing a novel network architecture with an impressive 152 layers. Unlike plain networks such as VGGNet and AlexNet, ResNet incorporates skip connections that allow information to bypass one or multiple layers, effectively mitigating the vanishing gradient problem. ResNet was inspired by VGGNet but includes 34 simpler layers alongside these skip connections. The use of bottleneck layers within ResNet helps maintain reasonable time complexity while achieving impressive performance. A 1x1 convolution layer is added to the initial and the final convolutional layers. This technique reduces the network parameters without reducing its functionality. In this paper, studies mostly have focused on ResNet50, ResNet101, and ResNet 152.

3) EfficientNetB

EfficientNet, a neural network architecture designed for computer vision and image processing, was introduced by Google AI researchers in 2019. It stands out for its remarkable compactness and computational efficiency. What makes it particularly popular is its adaptability in terms of depth, width, and resolution using compound coefficients. The base EfficientNet-B0 model incorporates inverted bottleneck residual blocks from MobileNetV2 and squeeze-and-excitation blocks. This network comes in eight versions, ranging from B0 to B7, with B0 and B3 being the most widely favored versions in recent studies. Notably, the simplest variant, EfficientNet-B0, boasts 5.3 million parameters.

4) Inception

The Inception network unlike many CNN architectures that suffered from the vanishing gradient problem due to stacked convolutional layers, the Inception network addressed this issue by incorporating filters of multiple sizes within the same level. This innovation not only boosted speed but also enhanced accuracy. Instead of increasing depth, the Inception network widens, as illustrated in Fig 7. The addition of 1x1 convolutions prior to the 3x3 and 5x5 convolutions helped reduce the number of channels, thereby lowering computational costs. The original Inception model, known as Inception-v1, is often referred to as GoogLeNet. Subsequent versions, including Inception-v2 to v4 and Inception-ResNet, represent iterative improvements over the previous models.

5) U-Net

U-Net is a popular CNN architecture primarily designed for segmentation tasks. It operates on a raw input image and generates pixel-wise segmentation outputs. The network propagates end to end through a sequence of steps, encompassing both downsampling and upsampling phases, as illustrated in Fig 7. The downsampling phase, or the contraction path, enhances the network's ability to identify the 'What' aspects of objects while downplaying the 'Where' details. In this stage The image that is provided as input goes through a pair of 3x3 convolution layers along with the ReLU activation function. Subsequently, a 2x2 max-pooling operation with a stride of 2 is applied, progressively reducing the spatial dimensions of the image feature map at each level until it reaches the 'bottommost' layer. The upsampling phase is known as expansion. employs the transpose convolution technique to increase the size of the image and generate a high-resolution segmentation map. At the final layer, a 1x1 convolutional layer is utilized to map feature vectors to specific classes. This pixel-by-pixel classification gives the U-net the ability to localize and distinguish object boundaries accurately.

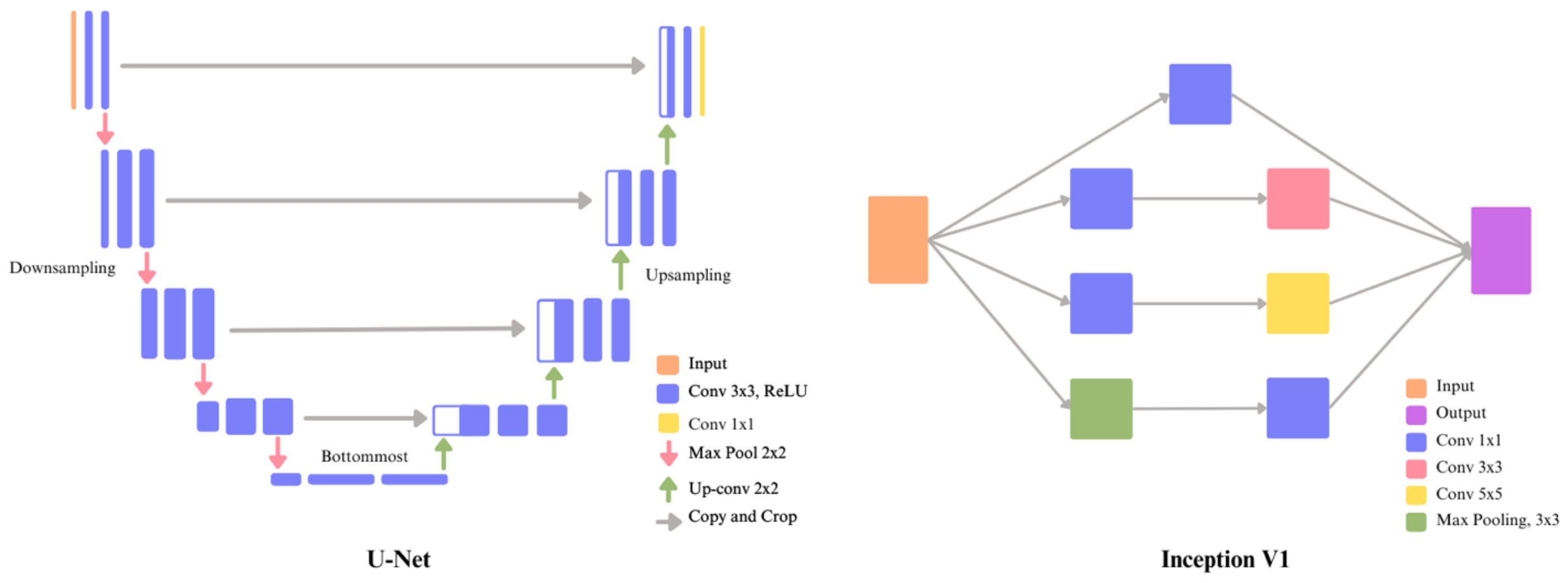

Fig. 7. Other popular deep learning architectures applied in cytology and colposcopy

## 4. Recent Deep Learning Methods in Cytology

Cervical cancer diagnosis relies significantly on the results of cytology screening. The precise differentiation between normal and abnormal cells constitutes a fundamental aspect of effective diagnosis. The identification of cancerous cells by pathologists is prone to errors and can be a time-consuming process. To address this issue, researchers have turned to deep learning techniques, making automated systems to improve cell detection and prediction accuracy, thereby contributing to the mitigation of diagnostic discrepancies. Additionally, the cytology view with alternations in cells, which are caused by HPV, is shown in Fig 8. Table 2 presents an organized overview of the information.

In cytology, various methods are employed, with Conventional Pap Smear (CPS) and Liquid-Based Cytology (LBC) being the most prevalent. According to [29], LBC tests are known for their effectiveness in producing satisfactory smears and accurately identifying both LSIL and HSIL. Additionally, the choice between CPS and LBC often depends on economic factors, with LBC being more commonly used in high-income countries, while CPS remains preferred in countries with lower to middle incomes [30]. Many cervical cancer screening programs are transitioning from CPS to LBC systems.

Two well-known LBC tests are SurePath and ThinPrep. [31] suggests that SurePath may result in higher rates of certain abnormalities, while ThinPrep tends not to significantly alter the cytological classification of smears or impact the rates of CIN detection. In the studies we reviewed, ThinPrep was the primary method used, although other methods were also employed.

The Bethesda System (TBS) is a standardized terminology and reporting system aimed at classifying Cervical Cytopathology results. It is designed to assist pathologists in categorizing the findings of cervical cytology specimens. Pap smear cells are typically divided into four main categories of epithelial cell abnormalities [32-33]:

1. Atypical Squamous Cells (ASC)
2. LSIL, which corresponds to CIN1
3. HSIL, which corresponds to CIN2 and CIN3
4. Squamous Cell Carcinoma (SCC)

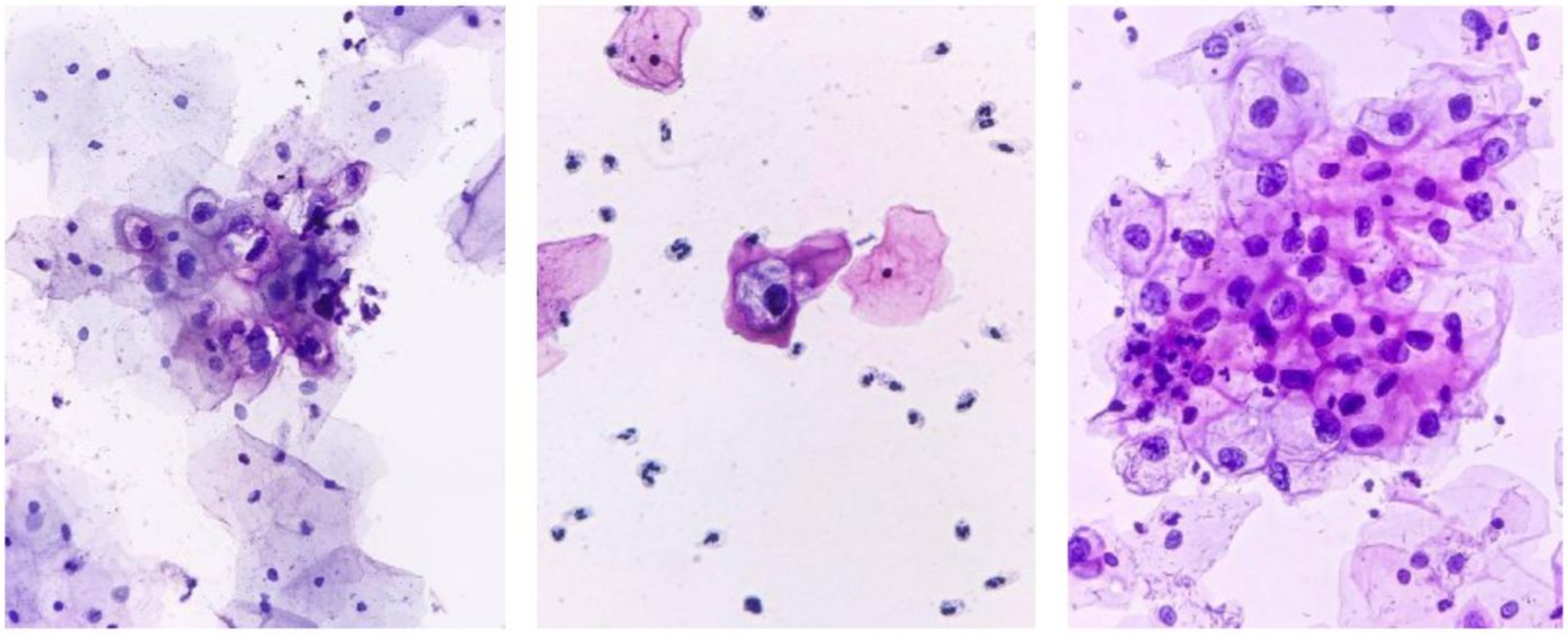

Fig. 8. Cytology view with cells changes, caused by the HPV virus

Wan et al. [34] adapted a DCNN for the segmentation of overlapped cervical cells in cytological images, employing a two-phase framework that included cell detection, cytoplasm segmentation, and boundary refinement. Their cell detection approach encompassed three key steps: nucleus candidate generation for image-wide classification, nucleus candidate selection to reduce false positives, and cell localization through a double-window approach for determining cellular regions. Their methodology involved using a pre-trained TernausNet model, derived from U-net, to classify pixels into the nucleus, cytoplasm, and background. Subsequently, they utilized a modified DeepLab V2 model to segment overlapping cytoplasm from the surrounding background and refine cell outer contours, aided by a novel loss function. Evaluation of three datasets (ISBI2014, ISBI2015, and an in-house private hospital dataset) and comparison with seven reference models yielded more promising results compared to the studies they compared with, particularly in boundary delineation and cellular segmentation. The process of transfer learning and creating pre-trained models is shown in Fig 9.

In a study conducted in 2020, Liu et al. [35] undertook the task of constructing a ThinPrep cytology dataset, sourced from four distinct hospitals. An automated classification system was developed capable of distinguishing between normal and abnormal cervical squamous cells. During the preprocessing stage, in accordance with the Bethesda reporting system, the images were labeled and resized to a uniform dimension of 224x224 pixels, guided by the bounding box coordinates. Subsequently, a DCNN model, based on the VGG16 architecture, was employed. To optimize model performance and robustness, an Ensemble Training Strategy (ETS) utilizing 5-fold cross-validation was adopted. The results were then compared with classifications made by two expert pathologists. the automated system achieved an accuracy rate of 98.07%, slightly more than the pathologist's performance while exhibiting significantly faster processing times.

In a recent study conducted by Albuquerque et al. [36], they introduced a groundbreaking non-parametric ordinal loss function for neural networks with the objective of guiding output probabilities toward a single-peaked distribution for the prediction of cervical cancer risk. The Harlev Dataset alongside different types of parametric and non-parametric loss functions to train nine distinct architectures, including AlexNet, GoogLeNet, and ResNet18, all pre-trained with the ImageNet dataset, was tested. Their research yielded accuracy rates of 75.6% and 81.3% for seven and four classes, respectively. They announced that in their experiment, the non-parametric losses outperformed parametric losses.

In Aliy et al. 's [37] paper, they conducted an evaluation of ten DCNNs for classification purposes. They employed transfer learning techniques when analyzing the pre-trained DCNNs on single-cell CPS images, initially pre-training these networks using the ImageNet classification dataset. For their analysis, they input preprocessed 128x128x3 images from SIPaKMeD into ten distinct feature extraction models, including ResNet, InceptionV3, and DenseNet among others. Subsequently, they developed a classification method for distinguishing among five different categories, with the most challenging cells to classify being Koilocytotic and metaplastic, which were initially often mistaken for each other.

A CYTOREADER was developed by Wentzensen et al. [38] , to enhance the efficiency and accuracy of CIN3+ detection in dual-slide images derived from ThinPrep and SurePath technologies. They utilized two neural networks: CNN4, consisting of 4 layers, and INCV3, comprising 48 layers, developed sequentially. In their experiments, CNN4 demonstrated strong performance on ThinPrep images, while INCV3 excelled with SurePath images. The system was trained and evaluated in three population studies, including cytology screening of HPV-positive women with SurePath samples. The automated evaluation of the DS-positive cells achieved an AUC (Area Under the Curve) of 0.82. They maintained that this platform is beneficial for women by reducing unnecessary colposcopies.

A DL-based WSI screening system for cervical cancer was constructed by Cheng et al. [39], combining Low-Resolution (LR) and High-Resolution (HR) models using ResNet50. The LR model pinpointed suspicious lesions, while the HR model identified lesion cells. The system recommended the top 10 lesion cells for pathologist review, and an RNN provided positive probabilities for WSIs. They used a private dataset with 12 groups of WSIs, including two independent evaluation groups. At the tile level, they achieved 95.3% specificity and 92.8% sensitivity, and at the slide level, 81.9% specificity and 79.3% sensitivity. Compared to the Hologic ThinPrep Imaging System, their system demonstrated a significantly higher TP rate due to its diverse image training.

A recent study by Yu et al. [40] addressed the issue of imbalanced sample numbers in real datasets, which often lead to low accuracy in classification tasks. They proposed a data augmentation strategy and conducted four tasks to assess its effectiveness. The results yielded an AUC value of 0.98 and an accuracy of 93.8%. Data utilized in this research was obtained from images of ThinPrep cytology tests gathered at a private hospital. To enhance their dataset, the authors employed a GAN model to generate additional abnormal cell image samples, which were then integrated with the existing private hospital dataset. They applied the CNN model AlexNet, which had been pre-trained on ImageNet, to classify normal and abnormal cells.

Another article that comprised both a CNN and a Recurrent Neural Network (RNN) was by Kanavati et al. [41], which trained the networks concurrently. The CNN component utilized the EfficientNetB0 architecture, with its outputs fed into the RNN model. The primary objective of this model was the classification of neoplastic and non-neoplastic specimens from LBC samples. The training dataset consisted of 1508 WSIs, while the validation set included 150 WSIs. Model evaluation was conducted on three distinct test sets: 'full agreement,' 'equal balance,' and 'clinical balance,' sourced from a private clinical laboratory. The model exhibited a strong performance, achieving a Receiver Operating Characteristic Area Under the Curve (ROC AUC) score of 0.96 on the test set.

More recently, new research has emerged on a DL approach for classifying WSIs in liquid-based cytology. The architecture of Alsalatie et al. [42] comprises CLS-Net, a CNN utilizing EfficientNet-B3 for region extraction, along with a pooling module known as Atrous Spatial Pyramid Pooling (ASPP) for capturing multiscale features. The process involved a two-step classification: first distinguishing between normal and abnormal cells, followed by further classification of abnormal cells into three classes (LSIL,

HSIL, SCC). They achieved an impressive accuracy of 99.6% when including preprocessing steps and 96.5% without preprocessing. Notably, their results surpassed those of other studies with which they conducted comparisons.

In the work of Xu et al. [43], they introduced another deep learning-based CNN for a detection task. The novelty in their approach was the use of COCO pre-trained weights for network initialization instead of the conventional ImageNet model, demonstrating the application of transfer learning. They conducted experiments with two architectures, ResNet50 and ResNet101, exploring different bounding box loss functions, models, and baselines. The results showed that ResNet50 outperformed ResNet101 as the backbone. They utilized Faster R-CNN with FPN detection algorithms as the baseline, achieving The SmoothL1 loss function delivers the most accurate model detection precision. To enhance performance, they applied fine-tuning and multi-scale training. Their experiments were conducted on a publicly available dataset containing 6,666 images for training and 744 images for testing, spanning across 11 categories. Ultimately, their model achieved an Average Recall (A) of 87.7% and a mAP of 61.6%.

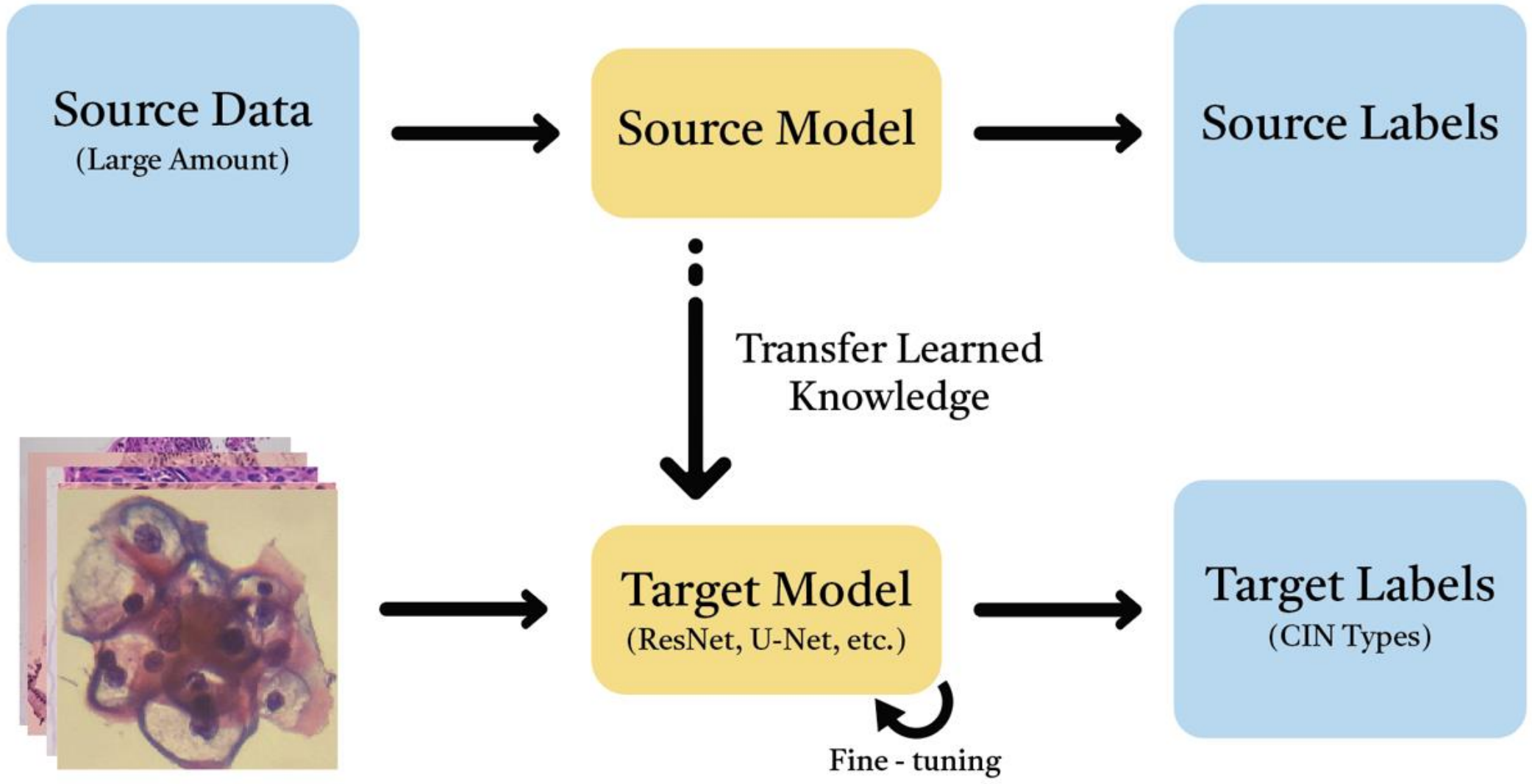

Fig. 9. How transfer learning helps to create models for cervical cancer diagnosis

In another object detection study, Xu et al. [44] developed a system for cytology using Papanicolaou (Pap) stained whole slide images. They collected a dataset of 500 WSIs from a hospital, resulting in a total of 5,721 pictures used for training and validation, with an additional 100 WSIs allocated for the test set. Their system employed Faster R-CNN as the base network structure, with ResNet101 used for feature extraction. To reduce FP, they applied data enhancement and result fusion techniques, along with foreground extraction, to limit computational load. The results of their experiment yielded 0.91 for precision in detecting positive cells on the validation set. For the test data, they encountered two-class and four-class problems with accuracies of 78% and 70%, respectively.

A two-step DL algorithm was developed by Nambu et al. [45] for the detection and classification of overlapped cervical cells in cytological images. The dataset, created by a medical university, adhered to the Bethesda system and underwent oversampling for improved learning accuracy. The algorithm initially employed YOLOv4 to detect atypical squamous cells, achieving an 85.7% average accuracy and a 54.2% precision across five atypical cell classes. Subsequently, ResNetSt, aided by a label smoothing algorithm, precisely classified the detected cells, boosting accuracy to 90.5% and precision to 71.8%. This approach holds promise for enhancing cervical cell analysis, particularly in liquid-based cytology applications.

Cervical Cell Copy-Pasting (C3P) was introduced by Stegmuller et al. [46] as an augmentation method that effectively transferred knowledge from open-source and single-cell datasets to unlabeled tiles, resulting in improvements for both tile-level and slide-level classification tasks. To alleviate the need for extensive manual annotation, the authors applied a self-supervised learning approach known as 'DINO,' which utilized a pair of Siamese teacher-student networks and a knowledge distillation technique during its pretraining phase. Additionally, the researchers created their own public dataset by compiling HPV-positive liquid-based cytology pap smear slides prepared using the SurePath method. To assess the performance of their models, evaluations were conducted at both the tile and cell levels, utilizing publicly available datasets like Sipakmed and Harlev. The experiments involved feature extraction using two distinct models, ViT-S/16 and ResNet-50. Notably, the inclusion of C3P yielded particularly favorable results, especially when ResNet-50 served as the backbone for the models.

In a recent study, Kurita et al. [47] established a DL approach to classify low-magnified cytology images with a less time-consuming labeling procedure. They utilized a private dataset from a hospital, comprising 140 WSIs from LBC specimens stained using the SurePath and standard Papanicolaou staining techniques, totaling 56,996 images split into training, validation, and testing sets. Initially, they trained an EfficientNet model using manually labeled images and generated additional pseudo labels for the unlabeled data. Subsequently, they fine-tuned it with 'noisy student training' which is a semi-supervised learning technique for binary classification into normal and abnormal categories, achieving an AUC of 0.90 with their test data.

Table 2. Information of cytology studies

| **Reference** | **Publish Year** | **Dataset** | **Preprocessing** | **Methods** | **Architecture** | **Cytology Methods** | **Results** |
|---|---|---|---|---|---|---|---|
| Wan et al. [34] | 2019 | ISBI2014, ISBI2015, an in-house private hospital dataset | - | Overlapping cell segmentation with two-phase DCNN | TernausNet and DeepLab V2 | - | ISBI2014: DSC 0.93 ISBI2015: DSC 0.92, In-house dataset: DSC 0.92 |
| Liu et al. [35] | 2020 | Multi-private-center dataset | Resized to 224*224 pixels, data normalization | A DCNN model with ETS to classify normal and abnormal cells | VGG-16 | ThinPrep | ACC 98.07% |
| Albuquerque et al. [36] | 2021 | Harlev | Resized to 224*224 pixels, zero padding, ImageNet similar normalization | Image classification into seven and four classes with non-parametric ordinal loss | AlexNet, GoogLeNet, MobileNet_V2, ResNEt18, ResNeXt50_32X4D ShuffleNet_V2_X1_0, | | seven classes ACC 75.6% four classes ACC 81.3% |
| | | | | | SqueezeNet1_0, VGG-16, Wide_ResNet50_2 | | |
| Aliy et al. [37] | 2021 | SIPaKMeD | - | Image classification into 5 categories | NASNetLarge, Xception, InceptionV3, ResNet101V2, ResNet101, InceptionResNetV2 ResNet152V2, DenseNet201, ResNet152, DenseNet169 | CPS | ACC 99% PREC 0.974 REC 0.974 F1-score 0.974 |
| Wentzensen et al. [38] | 2021 | Private dataset | - | A classifier consisted a series of 2 deep-learning models | CNN4 and Inception-v3 | ThinPrep and SurePath | AUC 0.82 |
| Cheng et al. [39] | 2021 | Private dataset from 5 hospitals | Data values are evenly spaced at a resolution of 0.243 micrometers per pixel. | Experimented two CNNs, Inception-ResNet-v2 and ResNet-152, for classification. | ResNet50 | ThinPrep | AUC 0.96, SPEC 93.5% SEN 95.1% |
| Yu et al. [40] | 2021 | Private dataset | Revised images to 227*227 | A GAN technology applied to images for CNN binary classification | AlexNet | ThinPrep | AUC 0.98 |
| Kanavati et al. [41] | 2022 | Private dataset | Used [1] method to eliminate the white background, | A CNN and RNN for neoplastic and non-neoplastic classification | EfficientNetB0 | ThinPrep | AUC 0.96 |
| Alsalatie et al. [42] | 2022 | Hussain et al. [26] dataset | A confusion matrix has been made for preprocessing and without it | Overlapped and non-overlapped cells are classified in 4 categories using CLS-N | EfficientNet-B3 | LBC | With preprocessing: ACC 96.5% Without preprocessing: ACC 99.6% |
| Xu et al. [43] | 2022 | Public dataset [58] | Normalization using the mean and std method | Faster R-CNN, FPN as baseline and RPN subsequently for ROIs selection and then classify them into 11 categories | ResNet1 01, ResNet50 | ThinPrep | AR 87.7%, mAP 61.6% |
| Xu et al. [44] | 2022 | Private dataset | The background area was removed, and foreground region cells were stained with Pap | A Faster-RCNN cell detection model with 2-class and 4-class classification | ResNet101 for feature extraction, RPN network for object prediction | Cervical cytology images stained with Pap | 2-class classification ACC 0.78, 4-class classification ACC 0.7 |
| Nambu et al. [45] | 2022 | Private dataset | Nine types of image filtering – scale augmentation and random erasing were performed on ResNetSt | Two-step detection and classification CNNs for cell clusters | YOLOv4 ResNeSt50 | LBC | ResNeSt ACC 90.5% and YOLOv4 ACC 85.7% |
| Stegmüller et al. [46] | 2023 | Trained with the introduced “in-house” dataset evaluated with Sipakmed and Harlev dataset | - | A self-supervised learning framework named as “DINO” for model pre-training | | SurePath | |
| Kurita et al. [47] | 2023 | Private dataset | - | Semi-supervised learning for labeled and unlabeled LPC images and a CNN model trained in purpose of binary classification | EfficientNet-B3 | SureAPth | AUC 0.910 |

## 5. Recent Deep Learning Methods in Colposcopy

The conventional colposcopy exam typically involves manual analysis of images and requires a high level of expertise. This can pose a challenge, particularly in low-middle-income countries. However, the use of DL technology has been found to be effective in accurately analyzing colposcopy images. This approach has the potential to reduce significantly the need for manual analysis and improve the precision of diagnoses. Table 3 provides a summary of relevant studies.

Zhang et al. [48] suggested a CAD system for the automation classification of cervical precancerous lesions using transfer learning and a pre-trained model named DenseNet CNN. First, they preprocessed the image data using ROI extraction to decrease the size of images and detect the cervix’s area. Data augmentation was also applied to improve the data. Then, they employed the parameters of all layers in a pre-trained DenseNet CNN from ImageNet and Kaggle. In addition, they examined the effects of several training approaches, such as random initialization (RI), fine-tuning a pre-existing model, various training data sizes, and implementing K-fold cross-validation. This method managed to acquire an accuracy of 73.08% and an AUC of approximately 0.75 in classifying CIN2 or higher-level lesions with 600 test images.

In an experiment by Yuan et al. [49], they utilized advanced DL models, including a classification model using multi-modal ResNet, a segmentation model using U-Net, and a detection model using Mask R-CNN. The models were fine-tuned using a pre-trained ResNet model, enabling them to capture cervical lesions and provide scientific diagnoses accurately. The study was performed on a large dataset of 22,330 cases, which included normal cases, LSIL cases, and HSIL cases. The input indices for the models had images of acetic and iodine and Information regarding patients' clinical details like their age, HPV testing and cytology results, and the type of transformation zone (TZ). The text information was coded using a specific method in the DL models. For instance, a person who is 45 years old and has tested positive for HR-HPV with an ASCUS cytology result and type 3 TZ would be assigned a coding of 01001010000001. The sensitivity and specificity of the classification model were 85.38% and 82.62%, respectively, successfully distinguishing between negative and positive cases. The model also achieved an AUC of 0.93. The segmentation model detected suspicious lesions with recall and Dice Coefficient (DICE) of 84.73% and 61.64% and an accuracy of 95.59%. The acetic detection model accurately detected high-grade lesions with a detection rate of 84.67%. The models' performance was better than colposcopists in ordinary colposcopy images. However, it had lower performance in high-definition images.

Moreover, Cho et al. [50] developed a model for accurately classifying colposcopic images, which can perform as well or better than experienced colposcopists. Two CNN architectures named the Inception-ResNet-v2 model and ResNet-152, were used for this matter. The pre-training of the CNN models was done using ImageNet weights, and then they were fine-tuned with the colposcopic images used in the study. Moreover, their method included a "Need-To-Biopsy System", which determined the need for a cervical biopsy. The AUC, on average, was 0.932 for accurately determining the need for a cervical biopsy and 0.947 for using the InceptionResnet-v2 and Resnet-152 models, respectively. Less experienced clinicians can benefit from this system as it assists in determining whether a cervical biopsy is necessary or if the patient should be referred to a specialist for further evaluation.

Using RestNet architecture, Lui et al. [51] employed ResNet architecture for the development of a CAD model using DL techniques. The choice of ResNet as the primary structure for the CAD system was motivated by its ability to effectively address the issues of vanishing and exploding gradients by introducing residual blocks. Furthermore, they devised a hybrid model that incorporates ResNet likelihoods and medical characteristics. The study involved a substantial dataset of 15,276 images, and the data obtained in the study indicated that the CAD system that incorporates both ResNet and clinical features results in a superior performance in comparison to the use of ResNet alone. These findings highlight the potential of DL techniques, such as ResNet architecture, for enhancing the accuracy and efficiency of CAD systems in the medical domain.

In addition to input data, Shinohara et al. [52] suggested a DL technique as diagnosis assistance in colposcopy, utilizing the U-Net model. This method utilized colposcopic images captured before and after the application of the acetic acid solution to segment lesions accurately. Both images were used as input data to train or test the U-Net network, which would output Segmentation Results. 30 real colposcopic images of acetowhite epithelium, which is a type of CIN cell, were used to test the method. As a pre-processing step, the images were aligned to better observe the alterations in each pixel when acetic acid was applied. This allowed for a more focused analysis of the changes in the images before and after the application. While some whitish areas of lesions were mistakenly interpreted as squamous epithelium in certain images, compared to using solely images captured after acetic acid application, the proposed technique yielded greater accuracy, precision, and F1 scores for segmentation. This study demonstrates that DL networks can benefit from the inclusion of images taken before and after acetic acid application to ensure accurate lesion segmentation.

Jiménez Gaona et al. [53] put forward a hybrid model for DL that utilizes both the U-Net network and machine learning techniques such as SVM. For the study, two sets of images were used. The initial set was sourced from the public dataset for Intel & Mobile ODT Cervical Cancer Screening. It comprised 1481 cervix images, which were further divided into normal and abnormal categories. The second set of images was gathered from females residing in a countryside community in Ecuador. The U-Net model is used for lesion segmentation, while SVM is implemented for classification. The U-Net model is trained using manually cropped ROIs extracted from colposcopy images. Data augmentation was employed to generate synthetic images from the ROIs, and both real and synthetic data were utilized to train the U-Net model for solving the segmentation issue. Following that, characteristics were extracted from the segmented ROIs, and classification was accomplished using SVM. The metrics for lesion segmentation show that the dice score is at 50%, precision is at 65%, and accuracy is at 80%. The classification results' sensitivity, specificity, and accuracy are 70%, 48.8%, and 58%, respectively [7]. It's important to acknowledge that the CAD system's capacity to anticipate cervical cancer risk needs improvement. However, it could be suitable in scenarios where women have restricted access to healthcare professionals.

Park and colleagues [54] employed a DL technique for unsupervised segmentation of the cervix area in colposcopy images. They utilized the Cervical Cancer Screening dataset from Intel & Mobile ODT. The W-Net architecture, which was adopted, comprises two U-Net architectures merged into a singular autoencoder. There are two components to U-Net: a decoder and an encoder. The input image undergoes compression into a smaller set of codes by the encoder, and then the decoder rebuilds the original image from the codes. As it undergoes training, the autoencoder becomes adept at compressing unmarked data into more streamlined code sets. The goal of the W-Net is to extract relevant traits from colposcopy images, which are used to segment the image into multiple classes. A method using normalized cut based on graphs is applied for the segmentation process, which groups pixels together according to their similarity. When the segmentation process was completed, the resulting regions were fed back into the model to train the model using a loss function called soft N-cut loss to generate a meaningful and accurate representation. Also, CT-loss and EW-learning were applied. CT-loss, a new combination of loss functions, improves the performance of the W-Net model. EW learning, which trains the encoder separately before training the entire W-Net, provides better convergence and segmentation accuracy. Throughout these processes, the suggested approach can achieve a DICE of 0.710, which reflects its good segmentation performance.

In another study, Kalbhor et al. [55] conducted a study where they tried out multiple feature extraction techniques in combination with different machine learning algorithms. They extracted features using a co-occurrence matrix (GLCM), a Gray-level run length matrix (GLRLM), and a histogram of gradients (HOG). Then, they put them as inputs into ML classifiers, such as Naïve Bayes, Bayes Net, Decision Table, Random Tree, Random Forest, and Logistic Regression. They also studied the classification while combining the features from different extractors, naming the hybrid feature fusion technique. Their dataset, which was gathered from WHO, was divided into two. Dataset-I had all kinds of images like aceto-whitening, green filter, Lugol's iodine, raw cervigram images, and Dataset-II with only aceto-whitening images being considered. As a result, for Dataset-I, the HOG feature extractor achieved a 69.72% accuracy, and for Dataset-II, the GLRLM feature extractor yielded the highest accuracy of 85.53%. The hybrid feature fusion technique improved the accuracies for both Datasets-I and II, reaching 72.43% and 84.47%, respectively. The data showed that the random forest classifier performed the best for both Datasets I and II. Moreover, the study confirmed that utilizing a hybrid feature fusion technique, combined with a random forest classifier, resulted in improved performance compared to other methods. The researchers emphasized that using acetowhitening images is essential for accurate cervical cancer detection and obtaining optimal results.

A novel neural network classification model using a hybrid algorithm has been developed by Chen et al. [56]. The model utilizes the EfficientNet-b0 architecture for extracting image features, and it conducts a fusion of three different types of examinations: original, acetic acid, and Schiller tests, using the GRU, which is a gated recurrency unit. The EfficientNet-b0 is a well-known CNN for its high parameter efficiency and processing speed. GRU is a type of RNN is known to enhance training efficiency considerably. There were 18,006 images collected during the research, which involved 612 individuals with HSIL, 1101 individuals with LSIL, and 4289 normal individuals. The novel model EfficientNet-B0 with GRU achieved higher accuracy (91.18%) compared to EfficientNet-B0 alone (88.32%) in categorizing normal controls, LSIL, and HSIL populations. This model was shown to be highly accurate and significantly better than other traditional neural network architectures, such as AlexNet and ResNet 50.

Another recent study was done by Yu and colleagues [57]. They introduced a model called CLS-Model that focuses on segmenting lesions in acetic acid colposcopic images. A Faster R-CNN is integrated into the CLS-model to accurately pinpoint the specific area, while also eliminating any potential interference from nearby tissues. The system includes a CLS-Net neural network that employs EfficientNet-B3 to extract features, as well as a modified ASPP module to capture multiple scale features. This experiment was studied on 5,455 CIN cases, and the CLS-model achieved impressive output in the segmentation of cervical lesions with accuracy, specificity, sensitivity, and dice coefficient of 93.04%, 96.00%, 74.78%, and 73.71%, respectively. In the proposed model, the CLS-Net surpassed other advanced segmentation techniques, such as U-Net, FCN8x, DeepLabV3, SegNet, and CCNet, in terms of metrics like DICE, Recall, Specificity, and Precision. Fig 10 illustrates the process outlined by this model.

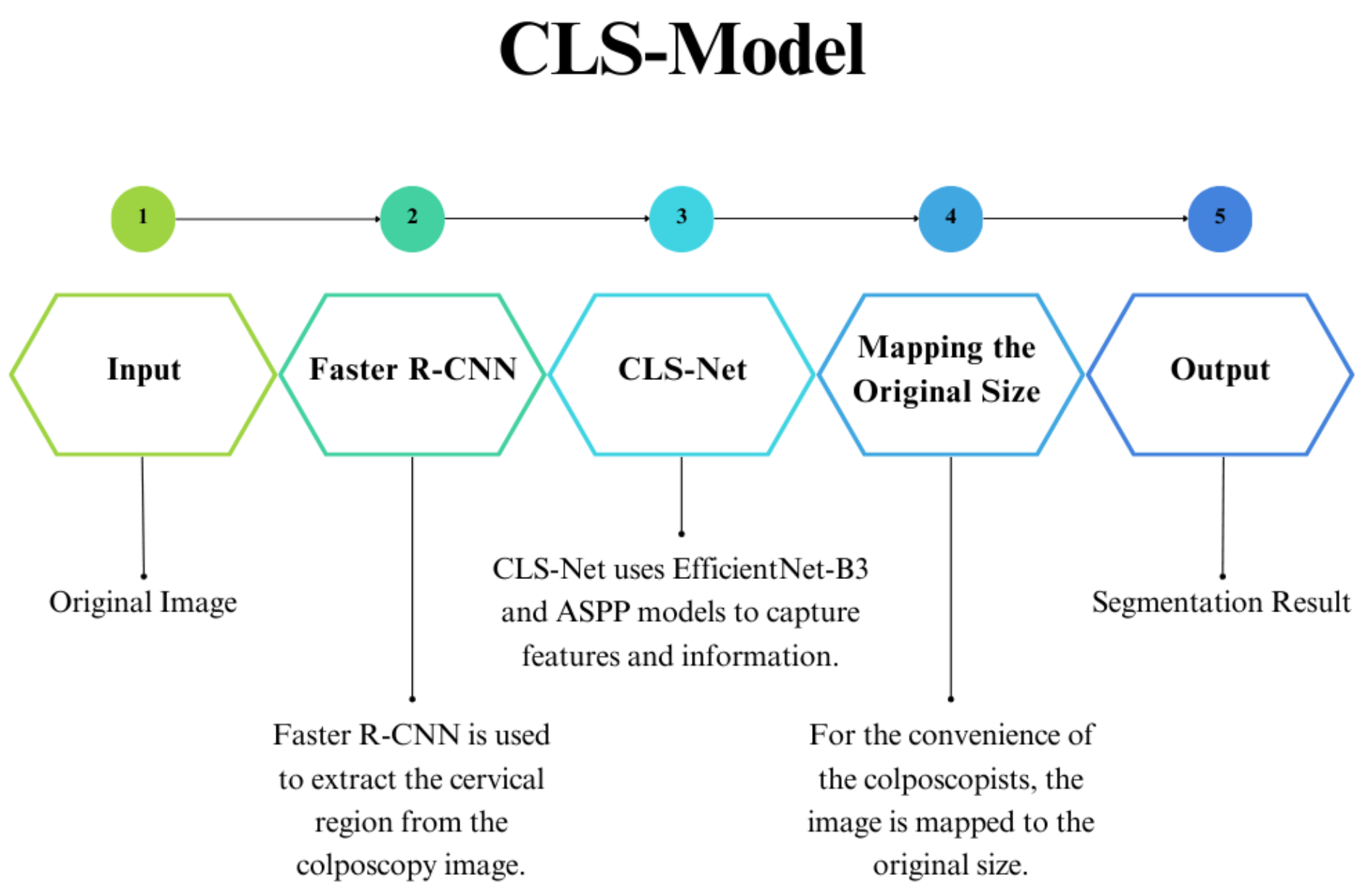

Fig. 10. Steps of segmenting colposcopy images in CLS-Model

Table 3. Information of colposcopy studies

| Reference | Publish Year | Dataset | Pre-Processing | Methods | Architecture | Results |
|---|---|---|---|---|---|---|
| Zhang et al. [48] | 2020 | Private dataset | The region of interest is extracted by experts. Different data augmentation methods are applied, such as Random Lighting Change, Random Blur, Random Crop, and Random Rotate. | Utilizing transfer learning and pre-trained DenseNet CNN for the CAD and classification | DenseNet | ACC 76.33% SEN 44.30% SPEC 87.78% |
| Yuan et al. [49] | 2020 | Private dataset | They resized one acetic and one iodine image per case to 512*512 pixels. Then, images were divided into 100 categories, and randomly separated into three sets: training, validation, and testing with the ratio of 8:1:1. Text information was coded. | A classification model that utilizes multi-modal ResNet, a segmentation model known as U-Net, and a detection model called Mask R-CNN. | ResNet, U-Net, and Mask R-CNN | Classification: SEN 85.38% SPEC 82.62% AUC 0.93 Segmentation: REC 84.73% DICE 61.64% ACC 95.59% |
| Cho et al. [50] | 2020 | A private dataset from 3 hospitals | The images were cropped to 480x480 pixels and normalized with min-max. Data augmentation was customized for each training dataset by adding rotated images. | Applying Inception-ResNet-v2 and ResNet-152 for classification | Inception-ResNet | Determining the need to biopsy: ACC 87.7% SEN 85.2% SPEC 88.2% |
| Lui et al. [51] | 2021 | Private dataset | Had a junior colposcopist outline ROI. Then, used data augmentation methods, including flipping/mirroring input images, rotating images, and shearing images. Finally, they resized all images to 224x224 pixels. | ResNet architecture and clinical features in CAD | ResNet | HSIL- vs HSIL+ CAD system: ACC 81% SEN 82.8% SPEC 80.3% |
| Shinohara et al. [52] | 2023 | Private dataset | A pre-processing step involved aligning the images captured before and after applying the acetic acid solution. | Segmentation by employing U-Net and analyzing the results before and after applying acetic acid. | U-Net | ACC 89.4% REC 85.5% PREC 83.7% |
| Jiménez Gaona et al. [53] | 2022 | Intel & Mobile ODT Cervical Cancer screening [20], and CAMIE (private) | Data Augmentation and the following transformations were used: contrast, brightness, gamma, rotation, and flipping the image horizontally. | A CAD system using U-Net for classification | U-Net | ACC 58% SEN 70% SPEC 48.8% |
| Park et al. [54] | 2022 | Intel & Mobile ODT Cervical Cancer screening [20] | No preprocessing was done. | An encoder-weighted W-Net with an additional pooling layer for segmentation of the cervix region. CT-loss and EW training are applied to the model. | W-Net | DICE 0.71 |
| Kalbhor et al. [56] | 2023 | Atlas of Colposcopy, WHO [21] | Pictures were adjusted to 100 × 200 and scaled down to 64 × 128. Data augmentation expanded the number of pictures. | Exploring hybrid feature fusion in ColpoClassifier for improved ML classification | ML classifiers | ACC 84.47% SEN 84% SPEC 18% PREC 84% |
| Chen et al. [57] | 2023 | Private dataset | The ROI was extracted by two experienced colposcopists, and then the images were adjusted to a size of 224 x 224 x 3. | EfficientNet-b0 as the backbone network for image feature extraction and GRU for feature fusion | EfficientNet | ACC 91.18% |
| Yu et al. [60] | 2022 | Private dataset | - | CLS-Model: A Segmentation method using Faster R-CNN and CLS-Net combined by EfficientNet-B3 and ASPP | Faster R-CNN, EfficientNet, and ASPP | ACC 93.04% SPEC 96.09 PREC 74.78% DICE 73.71% |

## 6. Discussion

Cervical cancer is one of the most common cancers and has high mortality rates among women. Therefore, early screening and diagnosis are essential for preventing the spread of the disease. Cervical screening can be done with various methods, including cytology and colposcopy. AI, especially machine learning and deep learning, has been shown to assist in cervical cancer image analysis, improving accuracy and efficiency. In addition, AI can provide a practical and affordable solution for low-income regions. In the diagnosis of cervical cancer, AI methods could be recommended as an alternative to inexperienced colposcopists and pathologists.

Various deep learning models have been trained to perform tasks such as image classification, segmentation, detection, and object localization. Preprocessing the cervical or cellular images as input data is a vital step. In most reviews, preprocessing is performed through Normalization, Data augmentation, and Fine-tuning. Among DL models, VGG-16, ResNet50 -101, EfficienNet-B0 -3, and Inception V3 are mostly used in Cytology, leading to a promising accuracy of 96%. In addition, for colposcopy ResNet, U-Net, and EfficienNet-B0 -3 are mostly applied, concluding an accuracy of 91% on average. These models improve the accuracy and efficiency of detection compared to traditional methods, which rely mainly on manual inspection. The advantages and drawbacks of each study are listed in Table 4.

Although AI-powered models have shown great potential in cytology and colposcopy, there are several challenges. Firstly, Open-source cervical screening dataset images are not sufficiently available. Secondly, the quality of data can be low, which might pose a major obstacle to the accuracy of the AI models. Thirdly, it is important to note that AI cannot replace clinicians and their expertise. Finally, the established AI-based models have not yet been widely adopted in clinical practice, and hence, prospective clinical studies are needed to verify their effectiveness.

In this review article, the importance of cervical screening, the role of AI in image analysis, different models for improving accuracy, and challenges related to AI implementation are discussed. The review article also presents the best accuracy in cytology, colposcopy, and the most used model. Additionally, preprocessing methods are highlighted. The paper concludes that future applications of AI might include not only early screening and diagnosis but also cancer treatment, prognosis prediction, and prevention.

Table 4. Advantages and Drawbacks of cytology and colposcopy studies

| **Cytology** | | | **Colposcopy** | | |
|---|---|---|---|---|---|
| **Reference** | **Advantages** | **Drawbacks** | **Reference** | **Advantages** | **Drawbacks** |
| Wan et al. [34] | Overlapped cytoplasm and nucleus segmentation | The cytoplasm and nucleus segmentation process were subsequently | Zhang et al. [48] | Using pre-trained models and transfer learning enables efficient training and fine-tuning of CAD systems, even with limited data. | Limited size of training data |
| Liu et al. [35] | Used different images of hospital databases | - | Yuan et al. [49] | It is based on both colposcopy images (acetic and iodine) and | - Does not provide details about the input text information coding methods. |
| | | | | clinical information. | -Has weaker output in high-definition images. |
| Albuquerque et al. [36] | Nine different architectures were compared alongside non-parametric and parametric loss functions. | - | Jiménez Gaona et al. [53] | Proves a high potential of cervical cancer diagnosis in limited settings. | - Low metrics<br>- Limited size of training data |
| Aliy et al. [37] | 10 different architectures were compared | Lack of generalization due to training and testing model with one dataset | Lui et al. [51] | Clinical features were included. | Improvement in the dataset is needed. |
| Wentzensen et al. [38] | Both tile level and slide level were conducted in the training process. | Other unrelated topics such as anal precancer detection were concluded. | Shinohara et al. [52] | - High precision, which minimizes sampling errors and physical burden on patients.<br>-Better recognition of appearance changes. | - Used a relatively small dataset of 30 actual colposcopic images.<br>-No significant difference compared to the controlled method (only with images after acetic acid application). |
| Cheng et al. [39] | Recommended the top 10 lesion cells to for pathology review, combining both low and high-resolution Walls. extensive dataset from 5 hospitals | - | Cho et al. [50] | The system included a "Need-To-Biopsy" feature, with superior performance from ResNet-152 | Models' metrics were low in classifying high-risk and low-risk lesions compared to other methods. |
| Yu et al. [40] | Employed GANs to address a few samples and class imbalance | Better architectures could be considered in classification models | Park et al. [54] | It can serve as a pre-processing step for automated CIN detection. | The information provided does not explain the amount of computational resources needed for training and inference. |
| Kanavati et al. [41] | Used different images of hospitals | - | Kalbhor et al. [56] | Improved accuracy can be achieved by utilizing a hybrid feature fusion vector | Used images from the WHO website, which could have limitations in terms of the quality and diversity |
| Alsalatie et al. [42] | The model had the ability to consider the slice of the image with overlapping and non-overlapping cervical cells instead of one single cell. | - | Chen et al. [57] | In this study, EfficientNet-B0 with GRU performed better compared to AlexNet and ResNet. | The basic clinical features need improvement for a high-quality study. |
| Xu et al. [43] | The promising performance of COCO pre-trained weights instead of the usual ImageNet. | - | Yu et al. [60] | - The results obtained were superior to those achieved by the most advanced techniques available.<br>- Can be visualized with a heatmap, which intuitively guides the detection of HSIL. | The reliance on only post-acetic-acid images |
| Xu et al. [44] | Counted the number of cells in each category, reduced FP with data enhancement and result fusion technique | Only ASC-US can be detected | | | |
| Nambu et al. [45] | All atypical cells can be detected and then classified. | Lack of generalization due to relying on one self-made dataset | | | |
| Stegmüller et al. [46] | Improved generalization across various modalities, including images that depict multiple cells | - | | | |

| | | |
|---|---|---|
| Kurita et al. [47] | Fewer labeled images are required for the training process. | - |

## 7. Conclusion

This review article explores various studies related to cervical cancer screening, which have been categorized into cytology and colposcopy. A total of fourteen cytology and ten colposcopy papers were reviewed for this study. The article also provides an overview of popular Deep Learning architectures and introduces public datasets for both cytology and colposcopy. Furthermore, this review can serve as a helpful resource for researchers interested in practical studies related to cervical cancer diagnosis.

### Abbreviations

| | |
|---|---|
| **HPV** | Human papillomavirus |
| **CIN** | Cervical Intraepithelial Neoplasia |
| **HSIL** | High-grade Squamous Intraepithelial Lesion |
| **LSIL** | Low-grade Squamous Intraepithelial Lesion |
| **VIA** | Visual Inspection of Acetic Acid |
| **PCR** | Polymerase Chain Reaction |
| **AI** | Artificial intelligence |
| **ML** | Machine Learning |
| **NLP** | Natural Language Processing |
| **DL** | Deep Learning |
| **ANN** | Artificial Neural Networks |
| **CNN** | Convolutional Neural Networks |
| **WSI** | Whole Slide Images |
| **CAD** | Computer-Aided Diagnosis |
| **GAN** | Generative Adversarial Networks |
| **H&E** | Hematoxylin and Eosin |
| **ROI** | Regions Of Interest |
| **DNN** | Deep Neural Network |
| **TP** | True Positive |
| **TN** | True Negative |
| **FP** | False Positive |
| **FN** | False Negative |
| **PREC** | Precision |
| **REC** | Recall |
| **SEN** | Sensitivity |
| **SPEC** | Specificity |
| **ACC** | Accuracy |
| **DSC** | Dice Similarity Co-efficient |
| **IoU** | Intersection over Union |
| **ZSI** | Zijdenbos Similarity Index |
| **mAP** | Mean Average Positive |
| **AR** | Average Recall |
| **DCNN** | Deep Convolutional Neural Network |
| **CPS** | Conventional Pap Smear |
| **LBC** | Liquid-Based Cytology |
| **TBS** | The Bethesda System |
| **FNA** | Fine-Needle Aspiration |
| **ASC** | Atypical Squamous Cells |
| **SCC** | Squamous Cell Carcinoma |
| **ETS** | Ensemble Training Strategy |
| **AUC** | Area Under the Curve |
| **RNN** | Recurrent Neural Network |
| **ROC-AUC** | Receiver Operating Characteristic Area Under the Curve |
| **ASPP** | Atrous Spatial Pyramid Pooling |

## Authors Biography

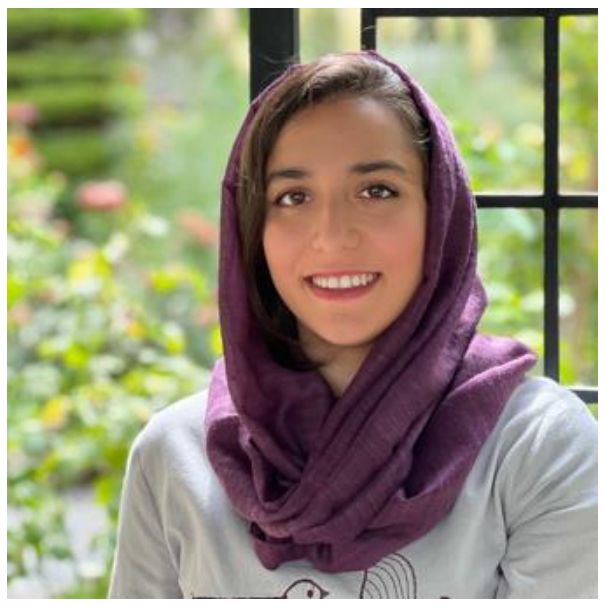

**Hana Ahmadzadeh Sarhangi**

BSc, Computer Engineering from Science and Research Branch, Islamic Azad University, Tehran, Iran.
Fields of Interest: Deep Learning, Computer Vision, Bioinformatics, and Software Development.

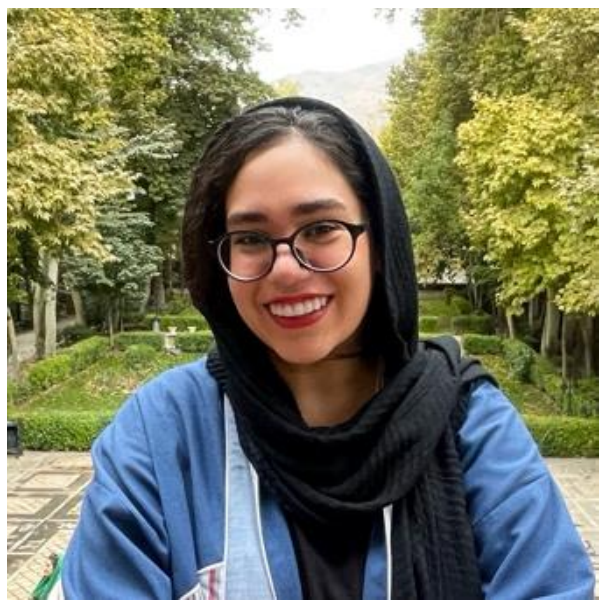

**Dorsa Beigifard**

BSc, Computer Engineering from Science and Research Branch, Islamic Azad University, Tehran, Iran.
Fields of Interest: Computer Engineering, Deep Learning, Computer Vision,
Data Science, Software Engineering, and Bioinformatics.

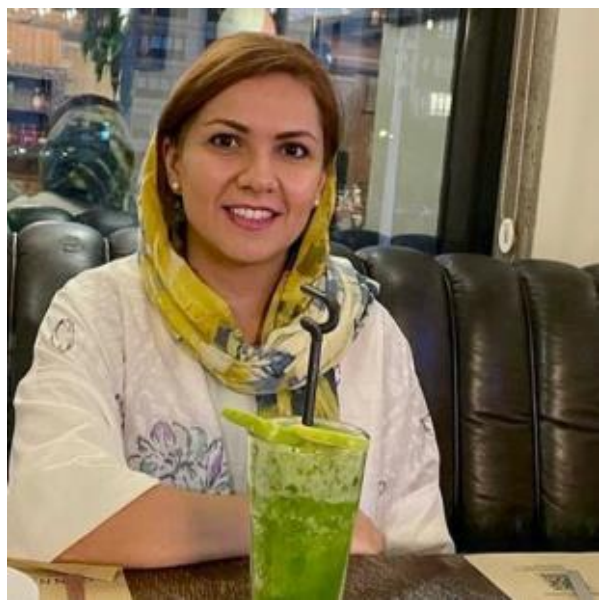

**Elahe Farmani, MD**

Anatomical and Clinical Pathologist
Tehran University of Medical Science, Tehran, Iran
Fields of Interest: Medical Science, Cancer Research, Molecular Pathology, Gynecology Pathology

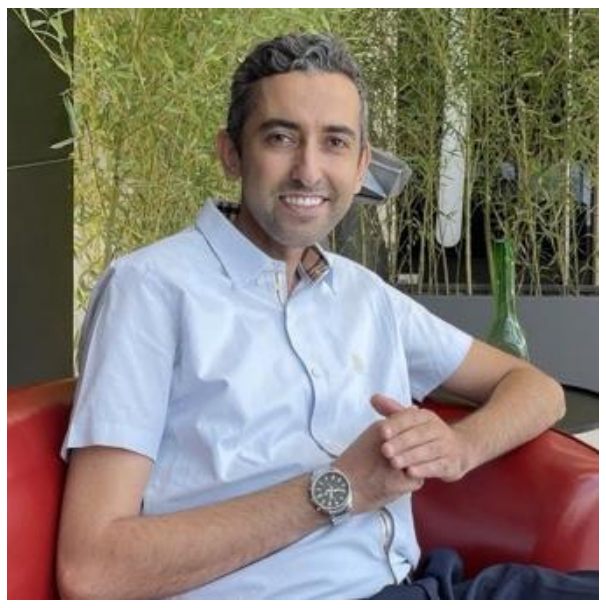

**Hamidreza Bolhasani, PhD**

AI/ML Researcher / Visiting Professor
Founder and Chief Data Scientist at DataBioX
Ph.D. Computer Engineering from Science and Research Branch, Islamic Azad University, Tehran, Iran. 2018-2023. / Fields of Interest: Machine Learning,
Deep Learning, Neural Networks, Computer Architecture, Bioinformatics